\documentclass[aps,pre,reprint,twocolumn]{revtex4-1}
\usepackage[pdftex,bookmarks,colorlinks]{hyperref}
\usepackage[pdftex]{graphicx}
\DeclareGraphicsExtensions{.pdf,.png,.jpg}
\usepackage{amssymb,ams fonts,amsmath,graphics}
\usepackage{pbox}
\usepackage[normalem]{ulem}

\begin{document}
\title{Drop impact experiments of non-Newtonian liquids on micro-structured surfaces}

\author{Marine Gu\'emas}
\author{Alvaro Marin}
\email{a.marin@utwente.nl}
\author{Detlef Lohse}
\email{d.lohse@utwente.nl}
\affiliation{Physics of Fluids, Faculty of Science and Technology, University of Twente}

\begin{abstract}
The spreading dynamics of Newtonian liquids have been extensively studied in hydrophilic and hydrophobic surfaces and its behavior has been extensively explored over the last years. However, drop impact of Non-Newtonian liquids still needs further study. Luu and Forterre (J. Fluid Mech., 632, 2009) successfully found scaling laws for yield-stress fluids on hydrophilic surfaces. They also uncovered interesting and yet unexplained regimes when the impact was performed on a superhydrophobic surface. In this work, we perform drop impact experiments on micro-patterned surfaces with two types of non-Newtonian liquids: one showing shear-thickening behavior and another one showing shear-thinning. Our results show that a typical shear-thickening liquid as cornstarch -at least a the relatively low concentration of 30\%w/w- spreads according to the scaling laws of Newtonian liquids, whereas visco-elastic liquids as Carbopol behave as predicted by Luu and Forterre for impact on hydrophilic surfaces, but show different scaling laws when they impact on superhydrophobic surfaces.
\end{abstract}

\maketitle

\section{Introduction}

Liquid drops impacting on solid and liquid surfaces have fascinated both scientists and photographers ever since the development of high-speed photography \cite{worthington1876forms, edgerton1954flash} and more recently high-speed video \cite{thoroddsen1998evolution, thoroddsen2008high}. Due to the improvement of these techniques, Clanet et al.\cite{clanet2004} were able to measure the spreading dynamics of water drops on superhydrophobic surfaces and extract scaling laws. Recent technical improvements have brought a new high-speed X-ray technique developed by Zhang et al.\cite{Zhang}, which has shown new details of drop pool-splashing. Xu et al. \cite{xu2005drop} pointed out the dramatic effect of the air underneath the droplet on its splash, which can be suppressed when the air is removed. Following this line of research, Reyssat et al.\cite{Reyssat2010} and Tsai et al.\cite{tsai2010langmuir, tsai2011microscopic}, have also shown how microscopic patterning at the surface can not only promote the splash but can also modify the preferential directions of splashing, therefore reshaping the canonical corona splash \cite{edgerton1954flash} and its fingers\cite{thoroddsen1998evolution}; we will refer to this phenomenon as ``directional splashing''.
Several models and numerical simulations have been tested to reproduce the complex physics of spreading and splashing of the drop\cite{attane2007energy, Brenner2010, Zaleski}. In spite of the substantial improvement in computation capabilities in recent years, there is nowadays no realistic simulation of a drop impacting on a solid surface at high velocity that compares well with experiments. Only scaling laws have been suggested, rationalizing the experimental observations.
Regarding the pre-splashing stage, several scaling laws have been proposed in the literature to account for the droplet spreading of a Newtonian droplet on a solid surface: assuming that the kinetical energy is all dissipated by viscous friction with the surface, a maximum spreading $D_{max}\sim D_oRe^{1/5}$ has been suggested, where $Re$ is Reynolds number defined as $Re={\rho U_o D_o}/{\mu}$, with $\rho$ the droplet density, $U_o$ its initial velocity, $D_o$ initial diameter and $\mu$ the liquid drop viscosity. For inviscid liquids, one may expect the kinetical energy to be totally transmitted into the spreading, which will have an energetic cost proportional to the liquid surface tension $\gamma$. Therefore one may expect a scaling with the Weber number $We={\rho D_o U_o^2}/{\gamma}$ as $D_{max}/D_o\sim We^{1/2}$. However, Clanet et al. \cite{clanet2004} found that the rate of spreading is much lower, namely as $D_{max}/D_o\sim We^{1/4}$, which they physically explain by balancing surface tension and inertia. They suggested that a significant amount of energy is invested in generating flow inside the drop, which was supported by their observations. In the following, we will employ the  classical spread factor $(D_{max}-D_o)/D_o$ for the experimental data, but we will use $D_{max}/D_o$ when performing scaling analysis for convenience.
This scenario changes substantially when non-Newtonian liquids are used instead, which are very common in practical applications as ink-jet printing\cite{inkjet2004}. Bartolo et al. \cite{bartolo2007dynamics} studied how the effect of minute amounts of polymer in the liquid can strongly alter the contact line dynamics and critically affect the liquid film retraction due to the dominance of elastic effects over capillarity. Luu and Forterre \cite{luu} systematically studied the impact and spreading of yield-stress fluids as polymer microgels and clay suspensions in both partially wettable and superhydrophobic surfaces. They showed that the spreading of polymer microgel liquids in smooth glass is solely determined by an elastic Mach number. However, when the spreading was tested on superhydrophobic sand, the spreading was much larger than that predicted on smooth glass; they referred to this regime as \emph{super-spreading}. Such phenomenon seems to appear due to the frictionless characteristics of the surface, but a systematic study on this regime is still lacking. 
On the other hand, shear-\emph{thickening} liquids as cornstarch present interesting and surprising behavior as persistent holes and fingerings\cite{Deegan2004holes,deegan2010holes}, or as the non-monotonic motion of solid balls found by Kann et al. \cite{Kann} inside a concentrated cornstarch solution, which the authors could explain with a minimal jamming model. Interestingly enough, classical shear thickening or linear viscoelastic models can not account for any of these mentioned phenomena. Another surprising effect is that found by Roch\'e et al.\cite{Roche}, namely the viscoelastic behavior of a cornstarch solution during the pinch-off of a shear-thickening liquid thread. Recently, some studies have tried to interpret the shear-thickening behavior of concentrated particle solutions as a jamming transition, triggered not only by a critical shear stress but also by confinement\cite{Bonn2008corn}.

The aim of the present work is to study drop impact of non-Newtonian liquids like yield-stress microgels as those employed by Luu and Forterre\cite{luu} and like shear-thickening liquids as cornstarch on micro-patterned superhydrophobic surfaces at Weber number values up to 3000. These surfaces have been shown to be extremely effective altering the splash dynamics at high Weber numbers. However, the super-spreading regime has been only observed in surfaces with random roughness (hydrophobic sand\cite{luu}). For this reason, one of the main questions to enquire here is on the possible enhancement of the super-spreading regime by using micropatterned superhydrophobic surfaces.
\begin{figure}[h]
\includegraphics[width=0.3\textwidth]{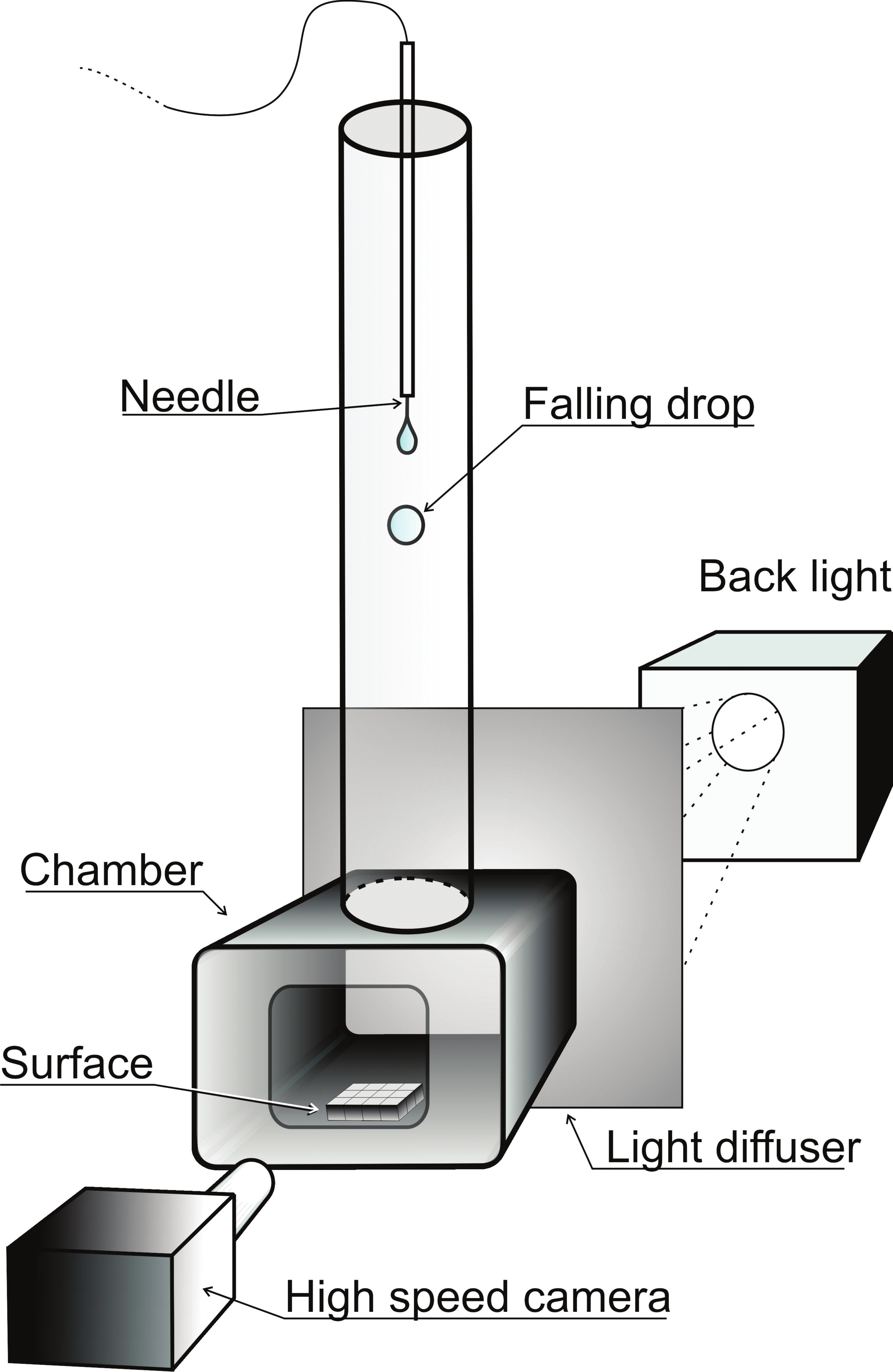}
\caption{Experimental setup employed for filming the side view and top view of impacting droplets.}
\label{setup}
\end{figure}

\section{Experimental}

The liquids employed for the first set of experiments consisted of the yield-stress polymer microgel \emph{Carbopol} (Carbopol ETD 2050, purchased from Sigma-Aldrich), prepared following the same recipe as Luu and Forterre\cite{luu}:~
De-ionized water was slowly added at 50$^\circ$ under constant mixing for two hours. At this point, the Carbopol solution was purchased and is highly acid ($pH=3$), and sodium hydroxide solution at 18\% is added to bring the pH up to 7. After stirring at 700 rpm to remove air bubbles and lumps, a transparent gel is obtained. Two different concentrations of Carbopol were used, 0.2\% and 1\%w/w. In the rheological measurements performed by Luu et al. \cite{luu} both Carbopol samples showed shear-thinning after a critical yield stress, with values $3.9$ and $24$ Pa respectively. They also were able to measure the elastic shear modulus of the samples by oscillatory measurements, yielding $28$ and $85$ Pa. The most relevant physical properties of the liquids employed have been summarized in table \ref{table0}.
For the second set of experiments, a cornstarch solution was prepared by mixing 30\%w/w of commercially available cornstarch with a density matched Cesium Chloride solution to avoid sedimentation.
In Fall et al.\cite{Bonn2008corn} the authors performed detailed rheological measurements and extended discussions on the origin of the peculiar rheological behavior of cornstarch. More concretely, within the range between 30-45\% w/w, the authors find a pronounced shear-thickening from a shear stress of approximately $20$ Pa.

\begin{table}[h]
\vspace{3mm}
\begin{tabular}[t]{|c | c | c | c | c |}
\hline

\pbox[t]{50mm}{ \textbf{Liquid}} &
\pbox[t]{70mm}{\textbf{~$\rho$($Kg/m^3$)~}}&
\pbox[t]{50mm}{\textbf{~G(Pa)~}} &
\pbox[t]{50mm}{\textbf{$~\sigma$($mN/m$)~}} &
\pbox[t]{50mm}{\textbf{$~\theta_c~$}} \\

\hline

\pbox[t]{50mm}{Carbopol 0.2\%} &
\pbox[t]{70mm}{1010} &
\pbox[t]{50mm}{28*} &
\pbox[t]{50mm}{70*} &
\pbox[t]{50mm}{$155^\circ \pm 10$}\\


\pbox[t]{50mm}{Carbopol 1\%} &
\pbox[t]{70mm}{1020} &
\pbox[t]{50mm}{85*} &
\pbox[t]{50mm}{70*} &
\pbox[t]{50mm}{$155^\circ \pm 10$}\\


\pbox[t]{50mm}{Cornstarch 30\%} &
\pbox[t]{70mm}{1200} &
\pbox[t]{50mm}{-} &
\pbox[t]{50mm}{50} &
\pbox[t]{50mm}{$160^\circ \pm 10$}\\

\hline
\end{tabular}
\centering
\caption{Properties of those liquid employed. $\rho$ is the liquid density, $G$ is the elastic modulus (only applicable for Carbopol), $\sigma$ is the surface tension and $\theta_c$ is the contact angle of a liquid sample drop measured on a structured PDMS sample (all pillared structures in table \ref{table1} gave indistinguishable results). Data marked with asterisk refers to data taken from Luu and Forterre\cite{luu}. All concentrations are given in weight percentage.}
\label{table0}
\vspace{3mm}
\end{table}

Regarding the wetting of the surfaces, Carbopol and cornstarch samples presented contact angles with the PDMS microstructure slightly smaller than those found with pure water and have been summarized in table \ref{table0}. 

The surfaces employed consisted on Polydimethylsiloxane (PDMS) micro-structures which have the advantage of being high reproducible and transparent. They are fabricated by etching the inverse pattern on a master replica mold on silicon. A degassed 10:1 w/w mixture of PDMS (Dow Corning-Sylgard Silicone Elastomer) base and curing agent is casted on the silicone mold and then cured in an oven where it is heated at 85$^\circ$C for three hours. Finally, the cured PDMS film is peeled off from the wafer and used. To examine the effect of the micro-pillar arrangement on the impact event, several molds were etched into cylindrical micro-holes with a certain depth \emph{h} arranged in a square or a hexagonal lattice. The dimensions of the final sample was $22.5 \times 22.5 mm^2$, which limited the size of our droplets to a maximum of 10mm in diameter. Hexagonal and square-lattice arrays of micro-pillars were used to explore the role of the geometrical distribution of micro-pillars of cylindrical micron-sized pillars of width \emph{w}, height  \emph{h}  and interspace  \emph{d}. In our experiments, \emph{w} is set to 5$\mu$m whereas \emph{d} and \emph{h} are varied respectively between 8, 15, and 25$\mu $m and 6, 10, and 20$\mu$m (see Table \ref{table1}). The liquids employed showed static contact angles in the range $150-170^\circ$ (see table \ref{table0}) and low hysteresis (limited to a few degrees). The results obtained from the impact on superhydrophobic surfaces will be contrasted with experiments on smooth untreated glass slides, which present partial wetting with the liquids used and static contact angles of $\sim 30^\circ$.

The aim of this work is to examine the spreading of non-Newtonian liquids on hydrophobic microstructured surfaces. However, some of the Carbopol samples were tested on hydrophobic sand for comparison with the work of Luu and Forterre\cite{luu}. The surfaces were prepared by carefully gluing a layer of hydrophobic grains on a glass slide. The grain size distribution is similar as that employed by Luu and Forterre \cite{luu}, with a highly polydisperse grain size distribution, with mean particle size $265\pm55$ $\mu m$. Contact angles of Carbopol drops on such substrates are slightly lower than those in PDMS micro-structured samples and presented much more dispersion.

\begin{table}[h]
\vspace{3mm}
\begin{tabular}[t]{|c|c|c|c|c|}
\hline
\pbox[t]{50mm}{\textbf{Sample}} &
\pbox[t]{50mm}{\textbf{d($\mu$m)}}&
\pbox[t]{50mm}{\textbf{h($\mu$m)}} &
\pbox[t]{50mm}{\textbf{Lattice}} &
\pbox[t]{50mm}{\textbf{Graph symbol}} \\
\hline
\hline
\pbox[t]{50mm}{a1} &
\pbox[t]{50mm}{8} &
\pbox[t]{50mm}{6} &
\pbox[t]{50mm}{Square} &
\pbox[t]{50mm}{\textcolor[rgb]{0.00,1.00,0.00}{$+$}}\\
\hline
\pbox[t]{50mm}{a2} &
\pbox[t]{50mm}{15} &
\pbox[t]{50mm}{6} &
\pbox[t]{50mm}{Square} &
\pbox[t]{50mm}{\textcolor[rgb]{0.00,1.00,0.00}{O}} \\
\hline
\pbox[t]{50mm}{a3} &
\pbox[t]{50mm}{25} &
\pbox[t]{50mm}{6} &
\pbox[t]{50mm}{Square} &
\pbox[t]{50mm}{\textcolor[rgb]{0.00,1.00,0.00}{$\Box$}} \\
\hline
\pbox[t]{50mm}{a4} &
\pbox[t]{50mm}{8} &
\pbox[t]{50mm}{6} &
\pbox[t]{50mm}{Hexagonal} &
\pbox[t]{50mm}{\textcolor[rgb]{0.00,1.00,0.00}{X}} \\
\hline
\pbox[t]{50mm}{a5} &
\pbox[t]{50mm}{15} &
\pbox[t]{50mm}{6} &
\pbox[t]{50mm}{Hexagonal} &
\pbox[t]{50mm}{\textcolor[rgb]{0.00,1.00,0.00}{$\triangle$}} \\
\hline
\pbox[t]{50mm}{a6} &
\pbox[t]{50mm}{25} &
\pbox[t]{50mm}{6} &
\pbox[t]{50mm}{Hexagonal} &
\pbox[t]{50mm}{\textcolor[rgb]{0.00,1.00,0.00}{$\diamondsuit$}} \\
\hline
\hline
\pbox[t]{50mm}{b1} &
\pbox[t]{50mm}{8} &
\pbox[t]{50mm}{10} &
\pbox[t]{50mm}{Square} &
\pbox[t]{50mm}{\textcolor[rgb]{0.00,0.00,1.00}{$+$}}\\
\hline
\pbox[t]{50mm}{b2} &
\pbox[t]{50mm}{15} &
\pbox[t]{50mm}{10} &
\pbox[t]{50mm}{Square} &
\pbox[t]{50mm}{\textcolor[rgb]{0.00,0.00,1.00}{O}} \\
\hline
\pbox[t]{50mm}{b3} &
\pbox[t]{50mm}{25} &
\pbox[t]{50mm}{10} &
\pbox[t]{50mm}{Square} &
\pbox[t]{50mm}{\textcolor[rgb]{0.00,0.00,1.00}{$\Box$}} \\
\hline
\pbox[t]{50mm}{b4} &
\pbox[t]{50mm}{8} &
\pbox[t]{50mm}{10} &
\pbox[t]{50mm}{Hexagonal} &
\pbox[t]{50mm}{\textcolor[rgb]{0.00,0.00,1.00}{X}} \\
\hline
\pbox[t]{50mm}{b5} &
\pbox[t]{50mm}{15} &
\pbox[t]{50mm}{10} &
\pbox[t]{50mm}{Hexagonal} &
\pbox[t]{50mm}{\textcolor[rgb]{0.00,0.00,1.00}{$\triangle$}} \\
\hline
\pbox[t]{50mm}{b6} &
\pbox[t]{50mm}{25} &
\pbox[t]{50mm}{10} &
\pbox[t]{50mm}{Hexagonal} &
\pbox[t]{50mm}{\textcolor[rgb]{0.00,0.00,1.00}{$\diamondsuit$}} \\
\hline
\hline
\pbox[t]{50mm}{c1} &
\pbox[t]{50mm}{8} &
\pbox[t]{50mm}{20} &
\pbox[t]{50mm}{Square} &
\pbox[t]{50mm}{\textcolor[rgb]{0.00,0.00,0.00}{$+$}}\\
\hline
\pbox[t]{50mm}{c2} &
\pbox[t]{50mm}{15} &
\pbox[t]{50mm}{20} &
\pbox[t]{50mm}{Square} &
\pbox[t]{50mm}{\textcolor[rgb]{0.00,0.00,0.00}{O}} \\
\hline
\pbox[t]{50mm}{c3} &
\pbox[t]{50mm}{25} &
\pbox[t]{50mm}{20} &
\pbox[t]{50mm}{Square} &
\pbox[t]{50mm}{\textcolor[rgb]{0.00,0.00,0.00}{$\Box$}} \\
\hline
\pbox[t]{50mm}{c4} &
\pbox[t]{50mm}{8} &
\pbox[t]{50mm}{20} &
\pbox[t]{50mm}{Hexagonal} &
\pbox[t]{50mm}{\textcolor[rgb]{0.00,0.00,0.00}{X}} \\
\hline
\pbox[t]{50mm}{c5} &
\pbox[t]{50mm}{15} &
\pbox[t]{50mm}{20} &
\pbox[t]{50mm}{Hexagonal} &
\pbox[t]{50mm}{\textcolor[rgb]{0.00,0.00,0.00}{$\triangle$}} \\
\hline
\pbox[t]{50mm}{c6} &
\pbox[t]{50mm}{25} &
\pbox[t]{50mm}{20} &
\pbox[t]{50mm}{Hexagonal} &
\pbox[t]{50mm}{\textcolor[rgb]{0.00,0.00,0.00}{$\diamondsuit$}} \\
\hline
\hline
\pbox[t]{50mm}{Glass} &
\pbox[t]{50mm}{} &
\pbox[t]{50mm}{} &
\pbox[t]{50mm}{} &
\pbox[t]{50mm}{\textcolor[rgb]{1.00,0.00 ,0.00}{$\rhd$}} \\
\hline
\end{tabular}
\centering
\caption{Geometric characteristics and graph symbols for the surfaces employed. \emph{d} is the micro-pillar diameter and \emph{h} its height expressed in micrometers.}
\label{table1}
\vspace{3mm}
\end{table}

\begin{figure}
\includegraphics[width=0.5\textwidth]{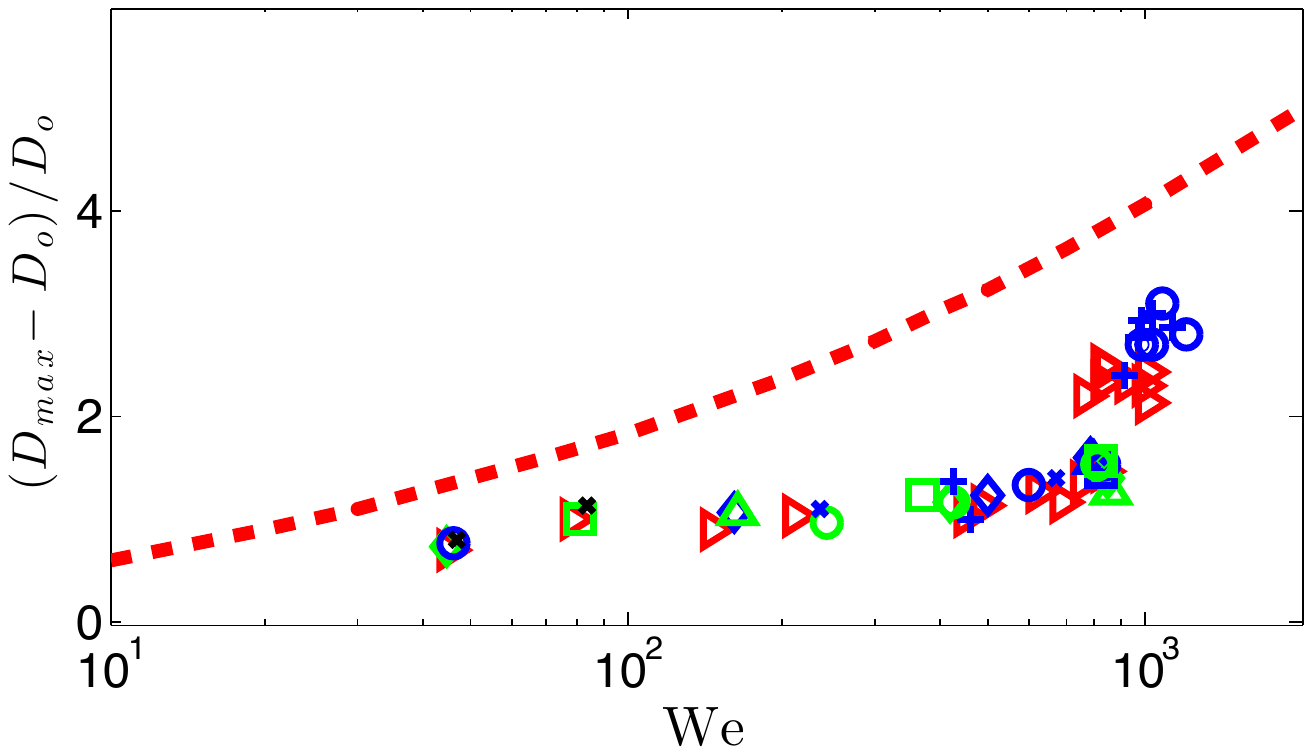}
\caption{The spread factor $(D_{max}-D_o)/{D_o}$ plotted as a function of Weber number for Carbopol 1\% drops. The scaling law found for water on superhydrophobic surfaces\cite{clanet2004}:$D_{max}/{D_o}=0.9 We^{1/4}$ is plotted as a dashed thick line as a reference.}
\label{plotCarbo1}
\end{figure}

The experimental setup for drop impact is sketched in figure \ref{setup}. Drops are formed in a capillary
needle with the aid of a syringe pump (PhD 2000, Harvard Apparatus) which feeds the drop until it falls by its own weight. Changing the needle outer diameter we were able to control the range of drop diameters from $\sim 2mm$ to 5mm. To vary the drop velocity, the drop was released from different heights, from 0.1 up to 3 meters, which resulted in velocities ranging from 1 to 8 m/s. A 500W spotlight illuminated the drop impact from the back through a light diffuser to create a homogeneous bright light. The drop impact is visualized with a high-speed camera (Fastcam SA1, Photron) with a recording rate ranging from 6000 to 10,000 fps (frame per second) and the aid of a long distance microscope (Edmund optics VZM300). Observation from the top is achieved with an inclined mirror mounted above the surface. Note that side view and top view are not recorded simultaneously to maximize the frame resolution.
\begin{figure*}
\includegraphics[width=\textwidth]{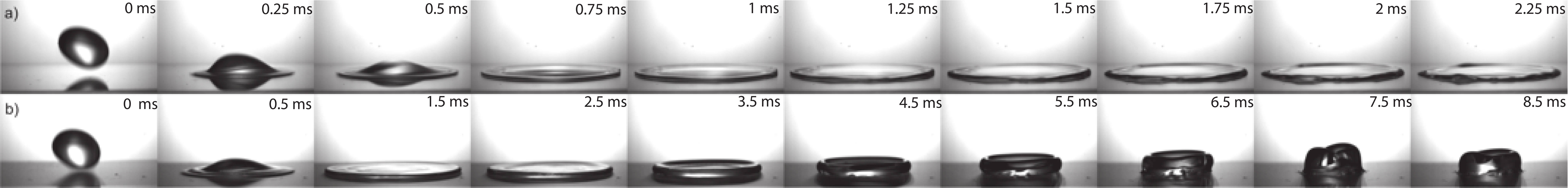}
\caption{Side-view sequence of snapshots of Carbopol 1\%tw drops of 4mm in diameter at 5 m/s ($We\sim1000$) (a) on smooth glass and (b) on a superhydrophobic surface. Note the lack of splash in both cases in spite of the high Weber number, and the fast recoiling of the droplet on the superhydrophobic surface.\label{sequence_carbopol1}}
\end{figure*}
\section{Results and discussion}

\subsection{Carbopol 1\%}

A typical impact event of a Carbopol 1\% drop on glass and on a superhydrophobic surface is shown in figure \ref{sequence_carbopol1}. In both cases, the drop deforms until a maximum diameter is reached. At this point, drops on glass remain attached to the surface, contrary to those on superhydrophobic surfaces, for which we observe a strong film recoiling. Particularly on the superhydrophobic surface, the lamella always spreads bending slightly upwards, as already observed and studied with both Newtonian \cite{xu2005drop,Brenner2010} and non-Newtonian liquids \cite{luu}. No partial or total rebounds have been observed with Carbopol 1\% on any of the surfaces due to the high Weber numbers regime which characterized this study. Due to the large liquid viscosity, fingerings or liquid crowns have not been observed either \cite{luu}.
In order to express the spreading of the liquid film in non-dimensional terms, we define the spread factor $(D_{max}-D_o)/{D_o}$ and plot it against the Weber number, ${We}=\rho D_o U_o^2/\gamma$, where $D_{max}$ is the maximum diameter that the film reaches, $D_o$ is the initial droplet size, $\rho$ is the liquid density, $U_o$ is the droplet impact velocity, and $\gamma$ is the surface tension of the droplet. In figure \ref{plotCarbo1} we plot the results for different droplets sizes, velocities and surfaces. As a reference, the scaling law found by Clanet et al. \cite{clanet2004} for water drops on superhydrophobic surfaces $D_{max}/D_o=0.9We^{1/4}$ is also plotted as a dashed line. The first thing one notes on the plot is that there does not seem to be a large difference between drop impacts on glass and in PDMS surfaces. Only around $We\sim1000$ some difference can be observed.
However, at this Weber number, Luu and Forterre \cite{luu} were able to appreciate a considerable larger spreading on their hydrophobic sand substrate than on glass. To contrast these results, we performed tests at $We \gtrsim 1000$ on hydrophobic sand, prepared by carefully gluing a layer of hydrophobic grains (Magic Sand TM) on a glass slide. The results were comparable to the results on the PDMS surface within our experimental error and therefore they are not plotted in figure \ref{plotCarbo1}.

\begin{figure*}
\includegraphics[width=\textwidth]{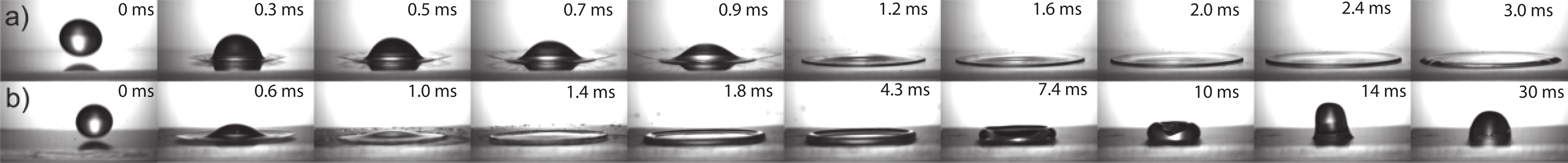}
\caption{Side-view sequence of snapshots of Carbopol 0.2\%tw drops of 4mm above the splashing limit (a) on smooth glass and (b) on a superhydrophobic surface. The impact on the superhydrophobic surface produces a strong recoiling, which occurs in the same time scale as in the 1\% case due to its capillary origin.}
\label{sequence_carbopol02}
\end{figure*}

\subsection{Carbopol 0.2\%}
In figure \ref{sequence_carbopol02} we show a sequence of images with an impact of drops of Carbopol 0.2\% on glass and on a PDMS surface. Unlike the more concentrated Carbopol solution, splashing impacts are observed on both glass and microstructured surfaces for similar Weber numbers. The drop behaves initially following the same behavior as described for Carbopol 1\%. However, above a critical Weber number, the thin lamella becomes unstable in the latest stage of the spreading and long fingers are created provoking the splash of the drop. Interestingly enough, no directional splashings \cite{tsai2011microscopic} have been observed for this particular sample even though water and ethanol droplets in this Weber number range do show a preferential direction for splashing. This might imply a strong dominance of visco-elastic forces over capillary ones in Carbopol solutions.
\begin{figure}
\includegraphics[width=0.5\textwidth]{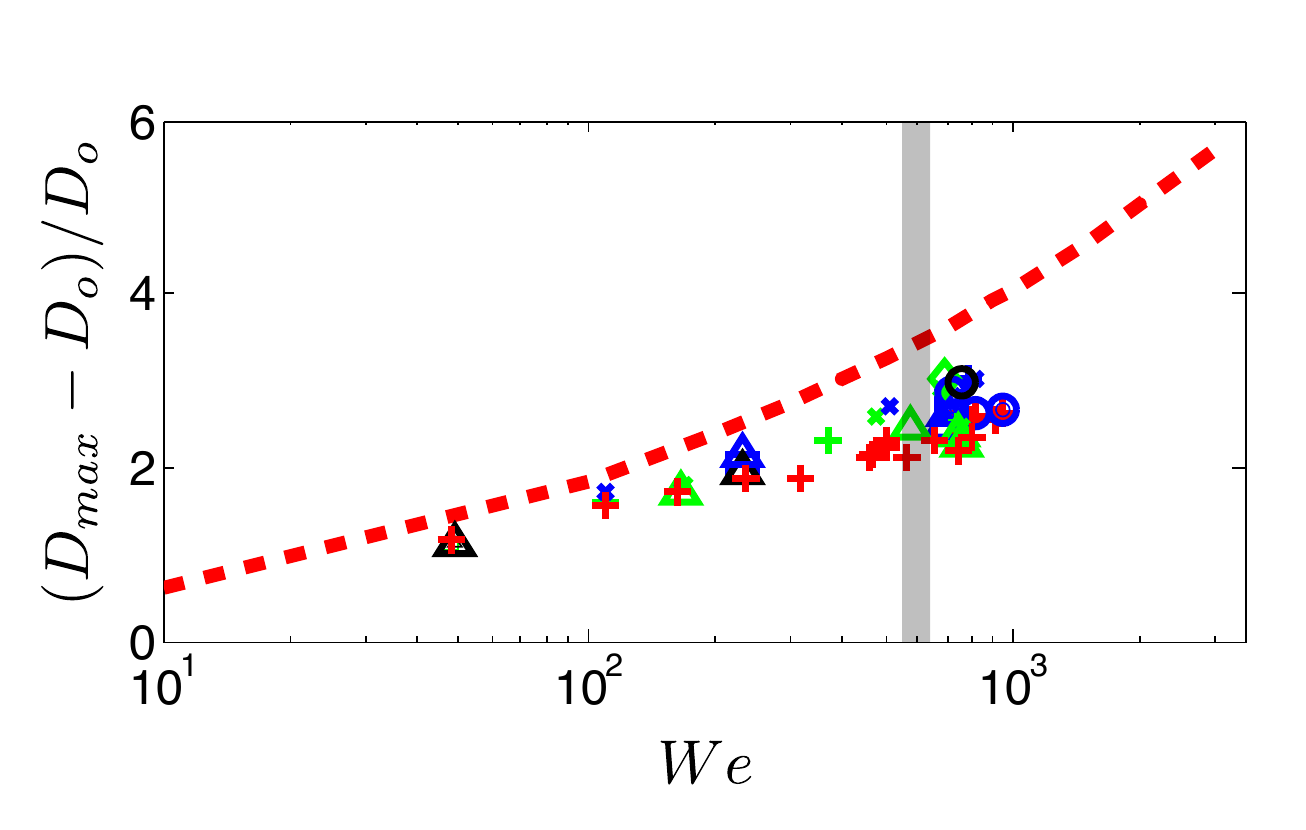}
\caption{The spread factor $(D_{max}-D_o)/{D_o}$ plotted as a function of Weber number for the solution of 0.2\% Carbopol. The scaling law found for water on superhydrophobic surfaces \cite{clanet2004} is plotted as a discontinuous thick line as a reference. The vertical grey line shows the range in which splashing is triggered.}
\label{plotCarbo02}
\end{figure}

On figure \ref{plotCarbo02}, the maximal spreading factor is plotted versus the Weber number. Again, only at high Weber numbers there seems to be some difference between the spreading on glass and on PDMS, but the effect is still much smaller than in the experiments of Luu and Forterre with hydrophobic sand \cite{luu}. As with carbopol 1\%, our attempts on hydrophobic sand gave analogous results as with micro-structured PDMS. In this case, splashing was found at $We\simeq570\pm50$ for PDMS structures and $We\simeq470\pm50$ on glass. Given the error in the measurements and the lack    

\begin{figure}
\includegraphics[width=0.5\textwidth]{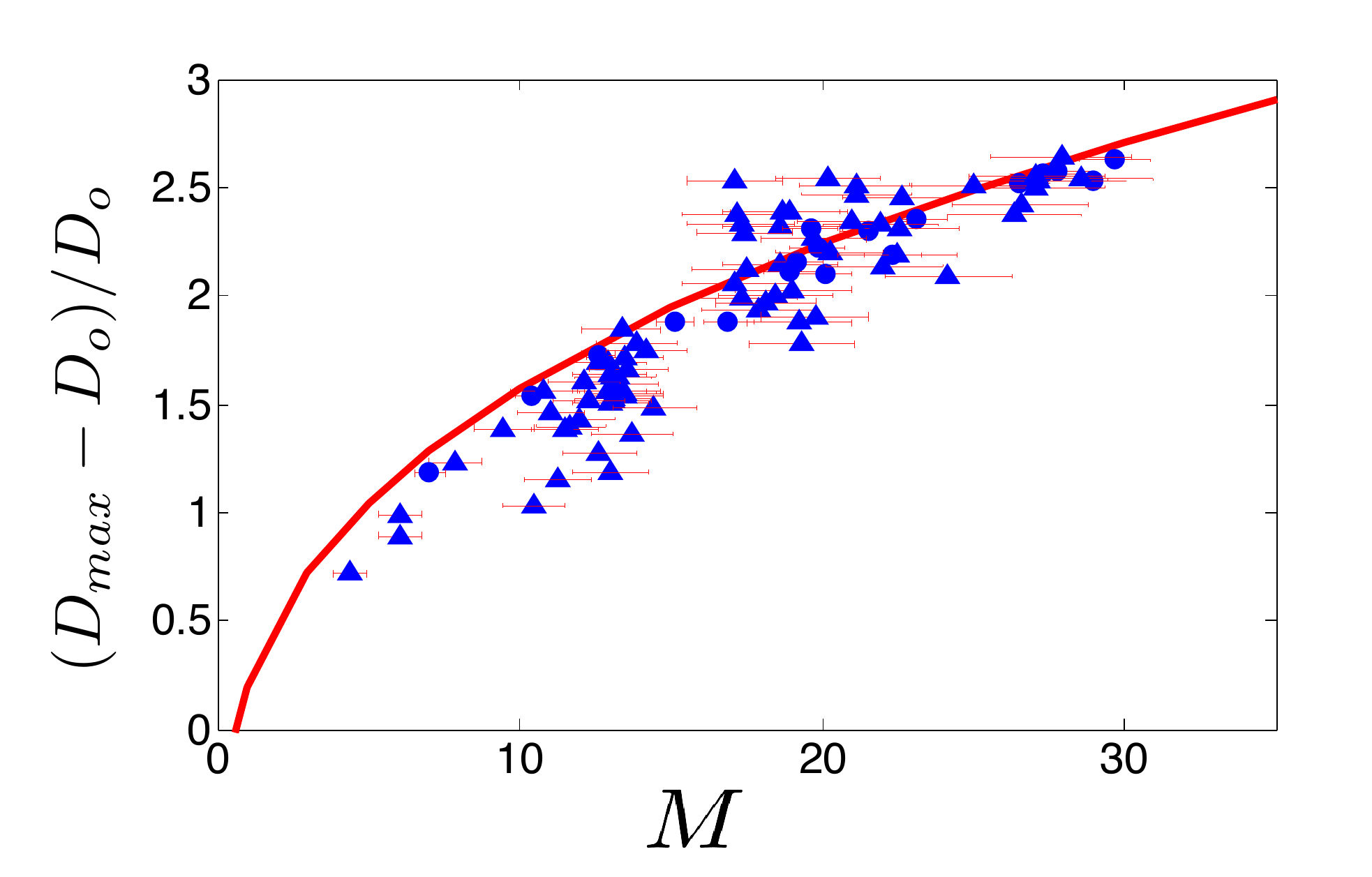}
\caption{The spreading factor $(D_{max}-D_o)/D_o$ from impact experiments on partially wetting glass, plotted as a function of the elastic Mach number M defined by Luu and Forterre\cite{luu}. The solid line corresponds to the scaling law by Luu and Forterre\cite{luu} for Carbopol solutions: $D_{max}/D_o \sim M^{1/3}$.}
\label{Machplot}
\end{figure}

Form this last result we must conclude that the super-spreading regime observed by Luu and Forterre \cite{luu} is not observed under our experimental conditions. One could argue that some differences in the preparation of the liquids might be the reason for the discrepancy. Although the liquid solutions were prepared following the same procedures and using the same concentrations, there may still be differences. 
To clarify this matter, we compared our results of Carbopol on flat glass with the scaling law proposed by Luu and Forterre\cite{luu} $D_{max}/D_o \sim M^{1/3}$, in which the elastic modulus $G$ of the liquids (see table \ref{table0})  is used to define an elastic Mach number $M=U_o/\sqrt{G/\rho}$, where $\sqrt{G/\rho}$ is the elastic velocity. The result is given in figure \ref{Machplot}, where we can see that our data fit reasonably well with their scaling law, except for some dispersion which can be explain by small differences in the Carbopol solution preparation.

The only parameter left that is significantly different from the experiments of Luu and Forterre\cite{luu} is the droplet size, which is in our experiments within a range of 2.5mm to 5mm and from 8mm to 27mm in their case. This suggest that the spreading of visco-elastic droplets on superhydrophobic surfaces would follow a different scaling law than that shown in figure \ref{Machplot} for partially-wetting surfaces, i.e. with a dependence on the drop size (which is totally absent in the elastic Mach number). Since our results on glass fits with Luu and Forterre\cite{luu}, it is unclear to us why on a partially wetting surface the visco-elastic droplets spread independently of the droplet size (at least within a decade).

\subsection{Cornstarch 30\%}
\begin{figure*}
\includegraphics[width=\textwidth]{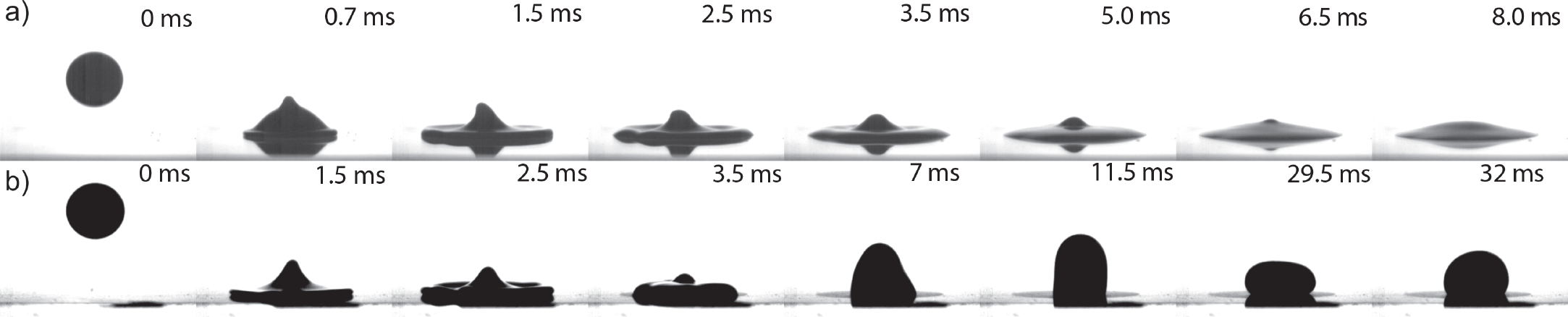}
\caption{Side-view sequence of snapshots of cornstarch suspension drop of 2.5mm at $U\simeq4.6m/s$ on (a) smooth glass and (total time 8.75ms)(b) superhydrophobic surface (total time 25ms). Note that the impact on the superhydrophic surface (b) does produce a significant splash as well as a strong recoiling.}
\label{sequence_maizena}
\end{figure*}

To investigate further the role of other non-Newtonian effects on drop impact dynamics, we perform drop impact experiments with a cornstarch suspension of 30\%w/w. A more careful study with a systematic variation of the cornstarch concentration will be done in future research. Unlike Carbopol solutions, cornstarch suspensions tend to thicken under high shear stress \cite{Bonn2008corn}, presenting opposite rheological properties as Carbopol. Figure \ref{sequence_maizena} depicts the different impact dynamics for a cornstarch drop on a PDMS micro-structured surface and on a smooth glass surface for an impact velocity of $U= 4.6m/s$. The first thing to note is that the cornstarch drop clearly splashes sooner on the micro-structured surface than on the smooth one. This phenomenon is common in Newtonian liquids \cite{yarin2006drop}, but is not very pronounced with Carbopol. Another fact we can observe in the sequence in figure \ref{sequence_maizena} is the strong recoil of the liquid film, which can also produce partial rebound for lower impact velocities.

\begin{figure*}
\includegraphics[width=\textwidth]{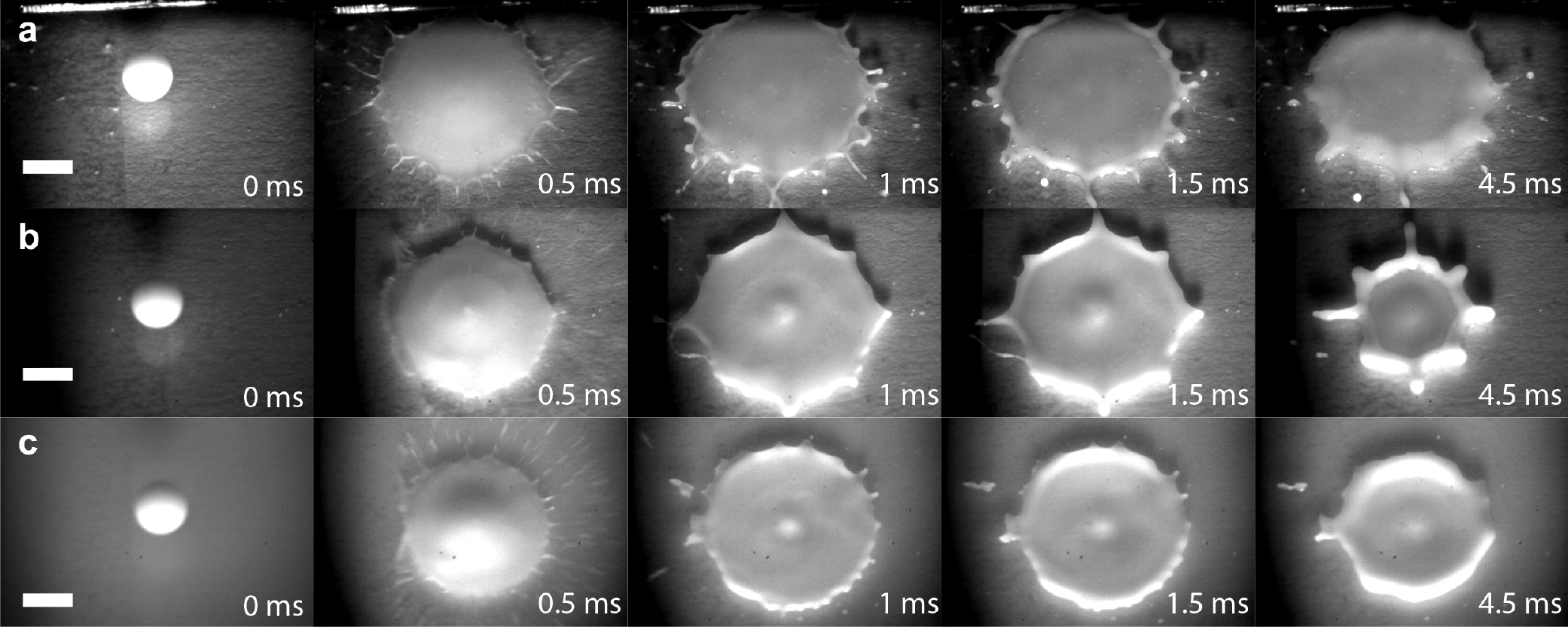}
\caption{Top view sequence of cornstarch drops impacting onto hydrophobic, square-lattice microstructures with three periodicities: (a) $d=8\mu m$, (b) $d=15\mu m$, (c) $d=25\mu m$. The bars indicate a length scale of 2mm.}
\label{top_maizena}
\end{figure*}

Furthermore, the geometrical arrangement of the hydrophobic micropillars does affect the cornstarch drop shape upon
splashing in a similar way as it does with Newtonian fluids \cite{tsai2011microscopic}. As can be seen in figures \ref{top_maizena}a and \ref{top_maizena}b, when the drop impacts on a distinct regular pattern, the splashing is driven in specific directions, which are strongly correlated with the lattice arrangement. In figure \ref{top_maizena} the pillar height \emph{h}, width \emph{w}, and drop's \emph{We} number are held constant for these experiments ($w = 5\mu m$, $h = 6 \mu m,$ and $We = 197\pm18$), and the lattice parameter \emph{d} the only varying one. The pillars are arranged in this case in square-lattice patterns. In contrast to the results of Reyssat et al.\cite{Reyssat2010} and Tsai et. al \cite{tsai2011microscopic}, the preferential directions are clearly observed not during the spreading and emission stage but during the \emph{recoiling} stage. The directional splashing is more clearly observed in cornstarch for a lattice parameter $d=15\mu m$ (Fig.\ref{top_maizena}b), similarly as reported by Tsai et al. \cite{tsai2011microscopic} for water, and almost suppressed for $d=25\mu m$ (Fig.\ref{top_maizena}c). The authors suggested that directional splashing must be originated by the air flow generated beneath the drop which is squeezed out through the empty spaces within the structure. This air flow might reach large enough velocities to drag some liquid with it and promote fingering and splashing formation \cite{Brenner2010} in the directions in which the liquid is allowed to flow, i.e. those directions permitted by the lattice arrangement. When the lattice parameter \emph{d} is big enough, air can flow at high enough velocities in additional directions and therefore the splashing pattern becomes more complex, as seen in figure \ref{top_maizena}c.

Interestingly enough, the cornstarch drops show a more pronounced \emph{directional recoiling} than spreading/splashing, which could be qualitatively understood in terms of the shear-thickening character of cornstarch: as the spreading occurs, shear stresses are strong and the effective viscosity of the film increases not homogeneously but in certain preferential directions, which oppose deformation and fingering formation. On the other hand, in the recoiling stage, the shear stresses relax and recoiling is driven by surface tension, which is then resisted by those lines which had been deformed and which present higher local viscosity. 

Besides this complex behavior and in spite of the expected local increase of viscosity, the 30\%w cornstarch drops seem to spread following the same scaling with the Weber number as a Newtonian fluid. This suggest that the shear stresses generated during a inertial splashing event (even at $We\sim1000$) does not seem to activate a significant non-Newtonian behavior at this cornstarch concentration. Roch\'e et al. \cite{Roche} found similar conclusions in a completely different configuration: they observed Newtonian behavior in the break-up of a cornstarch solution thread for concentrations below 32\%.

\begin{figure}
\includegraphics[width=0.5\textwidth]{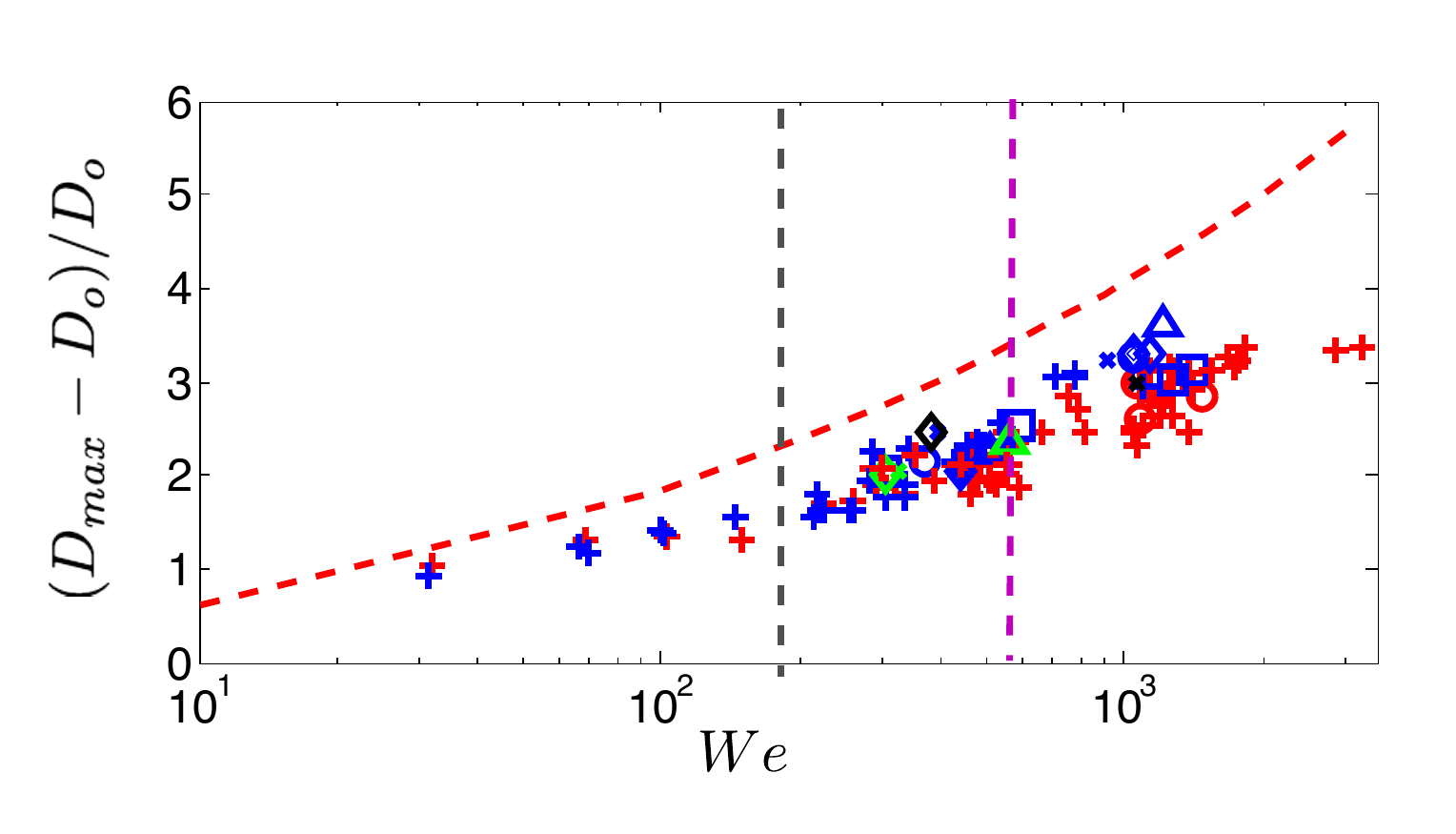}
\caption{The spreading factor $(D_{max}-D_o)/{D_o}$ for impacting cornstarch droplets plotted as a function of Weber number. The scaling law found for water on superhydrophobic surfaces \cite{clanet2004} is plotted as a dashed thick line as a reference. The vertical lines correspond to the splashing limits on PDMS (left) and on glass (right).}
\label{plot_maizena1}
\end{figure}

\begin{figure}
\includegraphics[width=0.5\textwidth]{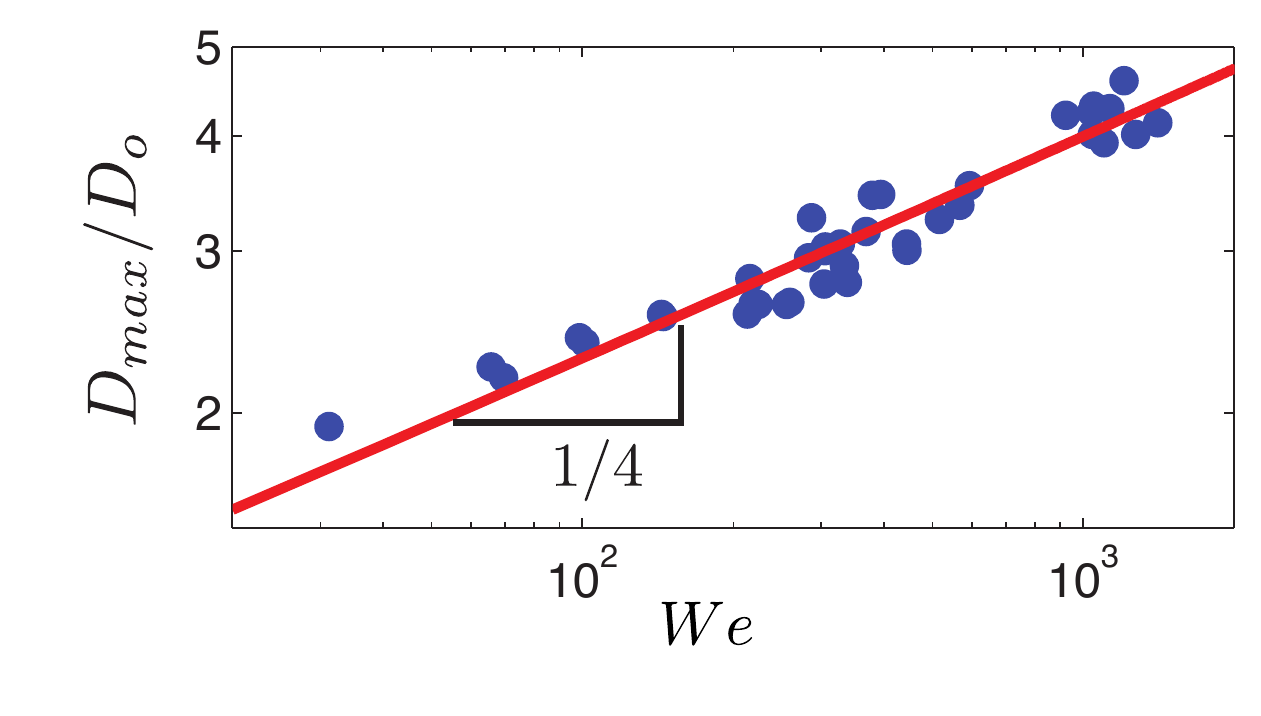}
\caption{$D_{max}/D_o$ for impacting cornstarch droplets plotted as a function of Weber number in logarithmic scale. The continuous thick line with slope 1/4 shows that cornstarch 30\% clearly follows the same exponent as that found for Newtonian liquids for Weber numbers up to 2000.}
\label{plot_maizena2}
\end{figure}

In figure \ref{plot_maizena1} we plot the spreading factor vs. the Weber number. Some difference is clearly observed between the spreading of cornstarch drops on glass and on PDMS microstructures at high Weber numbers. This is indeed a surprising effect since we expected that the super-spreading or the spreading enhancement was a phenomenon inherent to yield-stress fluids. Therefore, it might suggest that the phenomenon could be more general than expected. In figure \ref{plot_maizena1} it is also shown that the splashing is enhanced by the presence of the microstructure, which triggers the splashing at lower Weber numbers earlier than in glass. This fact has been already observed in previous studies \cite{luu, yarin2006drop, tsai2011microscopic} and could be related to the directional splashing and the preferential direction imposed to the air flow within the structure. On the other hand, in figure \ref{plot_maizena2} it is shown that cornstarch droplets in superhydrophobic surfaces clearly follow a 1/4 scaling law with the Weber number, as Clanet et al. \cite{clanet2004} demonstrated for inviscid Newtonian droplets. The data in figure \ref{plot_maizena2} follows a scaling law of the type $D_{max}/D_o=k We^{1/4}$, with a pre factor $k=0.497$ with a 90\% of confidence in our case. The prefactor $k$ is obviously smaller than that found for water by Clanet et al. \cite{clanet2004} ($k\simeq0.9$) due to the different liquid properties (mainly viscosity and surface tension).
This fact supports our previous conjecture about the Newtonian behavior of cornstarch at these relatively low concentration for these high Weber number splashing events, and might be supported by studies\cite{Bonn2008corn} suggesting that the shear-thickening character of cornstarch is a dilation effect which would vanish when confinement is absent as in our experiments.

Consequently, there is a remarkable difference with experiments as those from Kann et al.\cite{Kann}, in which a solid ball impacts a cornstarch solution at slightly higher concentrations (38\%w/w) and strong non-Newtonian effects are manifested. Apart from the higher cornstarch concentration, another relevant difference might be found on the way the particles in the liquid are driven by the flow: in Kann et al.\cite{Kann} the particles are suddenly compressed, which can promptly cause jamming when combined with confinement; while in our case they are pulled apart due to the liquid radial expansion in an unconfined system, which can hardly drive the system to jam. This can only be confirmed by performing further experiments with increasing cornstarch concentrations. According to this conjecture, the non-Newtonian behavior would appear at much larger concentrations than one would expect in a experiment with confined and compressing experiment.

\section{Conclusions}

In this work we have studied the different dynamics that arise when capillary drops of two different non-Newtonian liquids impact superhydrophobic surfaces at high velocities. In the case of visco-elastic liquids as Carbopol solutions, the liquid behave precisely as predicted by Luu and Forterre \cite{luu} when impacting into a smooth partially wetting surface. However, when the droplets impacted upon a micro-structured superhydrophobic surface the droplets did not show any extraordinary spreading as observed by the aforementioned authors. Since the initial droplet size is the only different parameter in our experiments, and since the scaling law proposed by Luu and Forterre is independent of it, we conclude that the impact dynamics of visco-elastic liquids as Carbopol on superhydrophobic surfaces must depend on the droplet diameter in a nontrivial way. The reason why the diameter does not enter into the scaling when the spreading occurs on flat hydrophilic surfaces is still unknown to us, but opens the door to a very rich phenomenology.
On the other hand, contrary to microgel \cite{luu} or polymer \cite{bartolo2007dynamics} solutions, which show a clear non-Newtonian behavior when only minute amounts of solute are added, cornstarch drops behave very different and at relatively high concentrations as 30\%w/w, they spread following the same scaling as that found with Newtonian fluids as water. With the given data we are not able to confirm whether this behavior is due to the low cornstarch concentration or due to the unconfined character of the experiment. If the latter were the case, it would confirm studies performed in the past which point out that the shear-thickening character of cornstarch is a dilation effect\cite{Bonn2008corn}, which would not manifest so easily when confinement is lacking in the system. However, we expect non-Newtonian effects to show up at higher cornstarch concentrations for impact experiments.
The more general conclusions of the present study is that there is a tremendously rich phenomenology of impact dynamics involving rheology and superhydrophobic microstructured surfaces, and it must yet to be discovered and explained in more detail.

The authors acknowledge Devaraj van der Meer, Stefan von Kann, and Matthieu Roch\'e for very fruitful discussions, and Peichun A. Tsai for the technical material suministrated.

\bibliographystyle{unsrt}

\end{document}